%% file: 00-main.tex
\documentclass[runningheads]{llncs}
\usepackage[utf8]{inputenc}
\usepackage[moderate]{savetrees}
\usepackage[T1]{fontenc}
\usepackage{xcolor}
\usepackage[inline]{enumitem}
\usepackage{wrapfig}
\usepackage{booktabs}
\usepackage{amsmath}
\usepackage{amsfonts}
\usepackage{amssymb}
\usepackage{enumitem}
\usepackage{mathbbol}
\usepackage{mathtools}
\usepackage{hyperref}
\usepackage{url}
\usepackage{nicefrac}
\usepackage{microtype}
\usepackage{natbib}
\usepackage{multirow}
\usepackage{graphicx}
\usepackage{subcaption}
\usepackage{etaremune}
\usepackage{balance}
\usepackage[font=scriptsize]{caption}
\usepackage{titlesec}

\newcommand{\eg}{\emph{e.g.}}

\titlespacing*{\section}{0pt}{0.5\baselineskip}{0.3\baselineskip}

\newif\ifanon
\anonfalse

\begin{document}
%\fancyhead{}

%\title{When do Humans vs Machines Benefit from the Full Text of Documents in Web Search?}
\title{Less is Less: When Are Snippets Insufficient for Human vs Machine Relevance Estimation?}
\titlerunning{When Are Snippets Insufficient for Human vs Machine Relevance Estimation}

\ifanon
\author{Anonymous\inst{1}}
\institute{Anonymous}
\else
\author{Gabriella Kazai \and
Bhaskar Mitra \and
Anlei Dong \and
Nick Craswell \and
Linjun Yang}
\institute{Microsoft, One Microsoft Way, Redmond, WA, USA \\ 
\email{\{gkazai,bmitra,anldong,nickcr,linjya\}@microsoft.com}}
\fi

\maketitle
\vspace{-1.5em}
\begin{abstract}
\input{01-abstract.tex}
\vspace{-0.8em}
\end{abstract}
\vspace{-1.5em}

\input{02-intro}
\input{03-related}

\input{04-design}

\input{05-method}
\input{06-results}

\input{07-conclusion}

\bibliographystyle{splncs04nat}
\bibliography{bibtex}

\end{document}

%% file: 01-abstract.tex
Traditional information retrieval (IR) ranking models process the full text of documents. Newer models based on Transformers, however, would incur a high computational cost when processing long texts, so typically use only snippets from the document instead. The model's input based on a document's URL, title, and snippet (UTS) is akin to the summaries that appear on a search engine results page (SERP) to help searchers decide which result to click.
This raises questions about when such summaries are sufficient for relevance estimation by the ranking model or the human assessor, and whether humans and machines benefit from the document's full text in similar ways. To answer these questions, we study human and neural model based relevance assessments on 12k query-documents sampled from Bing's search logs. We compare changes in the relevance assessments when only the document summaries and when the full text is also exposed to assessors, studying a range of query and document properties, e.g., query type, snippet length.
Our findings show that the full text is beneficial for humans and a BERT model for similar query and document types, e.g., tail, long queries. A closer look, however, reveals that humans and machines respond to the additional input in very different ways. Adding the full text can also hurt the ranker's performance, e.g., for navigational queries.  

%% file: 02-intro.tex
\section{Introduction}
\label{sec:intro}

In adhoc retrieval, ranking models typically process text from the URL, title and body of the documents. While the URL and title are short, the body may include thousands of terms.
Recently, Transformer-based ranking models have demonstrated significant improvements in retrieval effectiveness~\citep{lin2020pretrained}, but are notoriously memory and compute intensive. Their training and inference cost grows prohibitively with long input.
A common solution is to estimate document relevance based only on sub-parts of the document, e.g., query-biased snippets. Such approaches are motivated by the \emph{scope hypothesis}~\citep{robertson2009probabilistic}, which states that the relevance of a document can be inferred by considering only its most relevant parts.
Several neural approaches, \eg, \cite{hofstatter2021intra, yan2019idst}, have operationalized this hypothesis in their model design.
Document summaries based on URL, title and query-biased snippet (UTS) are also typically presented on SERPs to searchers. While the model uses UTS to estimate relevance when ranking, the human searcher uses UTS to estimate relevance when deciding whether to click a result. 
These scenarios motivate us to study when snippets are sufficient replacements of the full body text for relevance estimation by humans and machines. Concretely, by collecting human relevance assessments and relevance rankings from a machine-learned model both for UTS only and UTS plus body text inputs, we 
study whether humans and machines benefit from the document’s full text under similar conditions and in similar ways or if humans and machines respond to the additional input differently.  

%% file: 03-related.tex
\section{Related work}
\label{sec:related}
Automatic %extractive 
document summarization dates as far back as the foundational work by \citet{luhn1958automatic} and \citet{edmundson1964problems}.
In the context of search, several early user studies~\citep{tombros1998advantages, sanderson98tipster, white2003task} demonstrated the usefulness of query-biased snippets for assessing document relevance. % over query independent abstracts.
\citet{demeester2012snippets, demeester2013snippet} studied how well the document's relevance can be predicted based on the snippet alone in federated search. Unlike these prior works, our goal is to study the differences in human and machine relevance assessments when only document summaries or when also the body texts are inspected.
Past studies have also employed diverse measures of snippet quality based on manual assessment~\citep{kaisser2008improving}, eye-tracking studies~\citep{lagun2012re, cutrell2007you}, view-port analysis~\citep{lagun2011viewser}, historical clickthrough data~\citep{clarke2007influence, yue2010beyond}, and A/B testing~\citep{savenkov2011search}, but did not try to understand when and why human and model assessments differ.
%Other previous work have focused on improving snippet quality~\citep{li2008extracting, yulianti2017document}, but this is out of scope in the context of our current work.

The application of passage-based document views for adhoc document ranking have been explored in the context of traditional retrieval methods~\citep{bendersky2008utilizing, salton1993approaches}, but gained more attention recently~\citep{DBLP:journals/corr/abs-1901-04085, yan2020idst, Hofstaetter2020_sigir, hofstatter2021intra, li2020parade} in the context of Transformer-based~\citep{vaswani2017attention} neural ranking models.
While these models typically evaluate several passages per document, single query-biased summaries can be applied under stricter efficiency concerns.
Our work helps to understand the feasibility of estimating document relevance based on just the UTS information. 

Finally, our work is similar to \cite{bolotova-etal-2020} in the sense that we too study humans and a BERT model, but while \cite{bolotova-etal-2020} focused on attention, we study changes in relevance estimation due to input change. %our goal is to understand how humans and neural model change their UTS based relevance estimates due to the body text.   
%\citet{white2002finding}

%% file: 04-design.tex
\section{Experiment design}
\label{sec:design}

To answer our research questions, we collect both human and neural model based relevance assessments in two conditions: 1) when the human/machine assessor is only shown the query-biased summary, made up of the URL, title and snippet (UTS), and 2) when the body text is also exposed (UTSB). We use snippets returned by Bing's API.

\begin{figure}
  \centering
  \includegraphics[width=0.9\columnwidth]{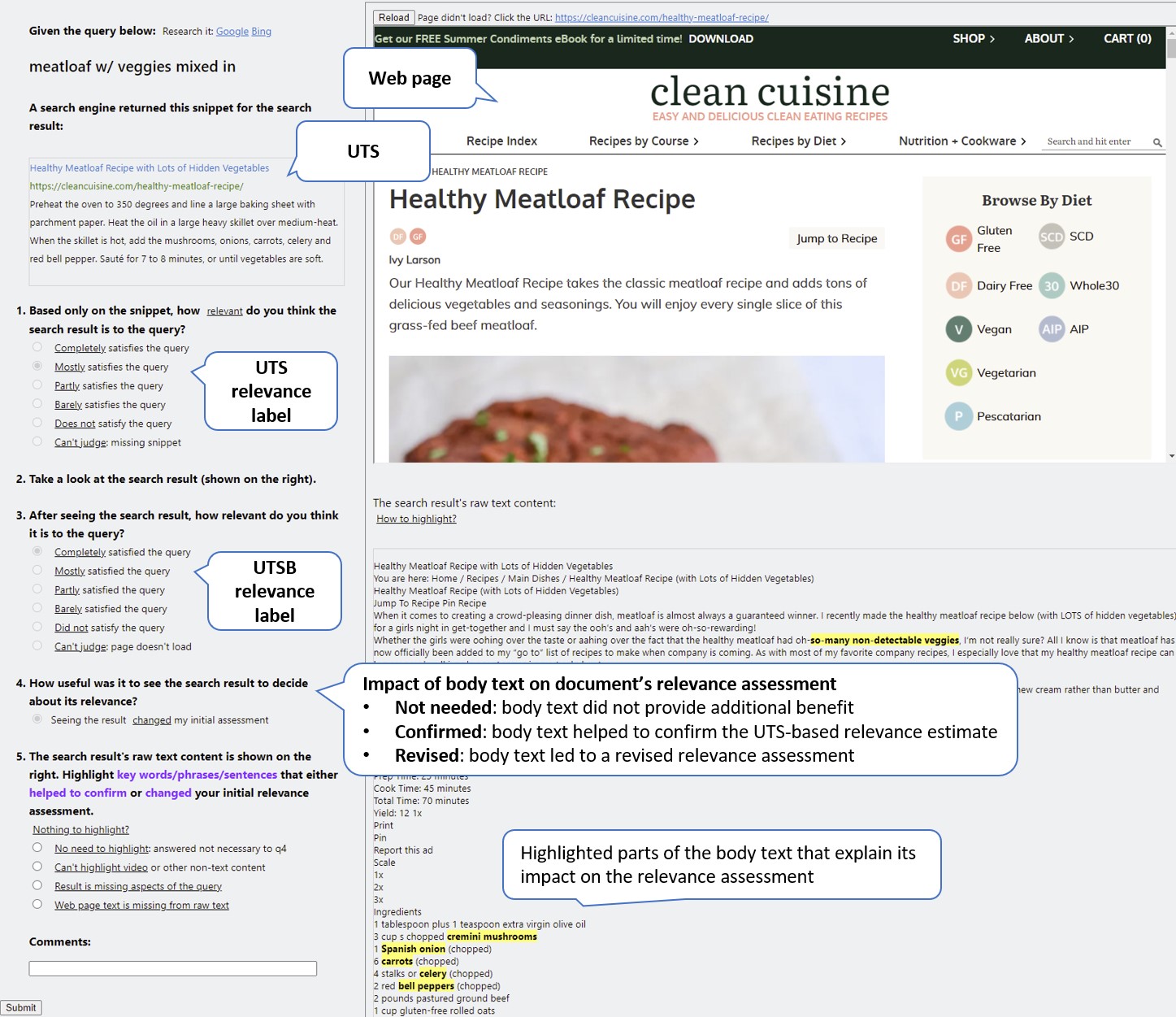}
  \caption{Human Intelligent Task to collect UTS and UTSB labels from assessors}
  \label{fig:hitui}
\end{figure}

We collect relevance assessments from humans via a Human Intelligent Task (HIT) with multiple judging steps, ensuring that the same person labels both conditions. First, we ask assessors to estimate a search result's relevance to the query based on its UTS information alone (UTS label). We then show assessors the web page and ask them to re-assess its relevance (UTSB label). Both labels use a five point scale.
Next, we ask if seeing the web page led to a revised assessment (`Revised'; this is auto-filled), if it helped to confirm the UTS based estimate (`Confirmed') or if the page did not provide further help in the assessment (`Not needed'). Finally, assessors are asked to highlight parts of the body text that explain why the body text provided additional benefit over the UTS. %the difference between their UTS and UTSB labels. 
Figure~\ref{fig:hitui} shows the final HIT state.
We use UHRS, an internal crowdsourcing platform, to collect judgments from trusted, quality monitored, long-term judges and pay them their standard hourly rate. We obtain an inter-assessor agreement rate of 0.44 for the UTS and 0.53 for the UTSB labels (Krippendorff $\alpha$).

%To train the models, we follow the standard learning-to-rank (LTR) practices: we prepared a set of <query, document> pairs whose relevance labels are available, and we split them into training set and validation set. Note that the relevance labels are provided by human judges who looked into the full body text as well as UTS texts to make their relevance labeling decisions. 

For our Neural Ranker based relevance estimation, we follow the state-of-the-art neural ranking approach \citep{DBLP:journals/corr/abs-1901-04085} and train a UTS and a UTSB ranker, starting with a pretrained BERT-style \citep{devlin-etal-2019-bert} model. 
The model inputs comprise sentence A, which is the query, and sentence B, which is either UTS or UTSB, respectively.
Query and UTS have an expected length of less than 128 tokens, so we use an input sequence length of 512 tokens in all our experiments, truncating the input if it is longer. 
This allows the UTSB model to see significantly more document text than is seen from snippet alone, and allows us to observe systematic differences between UTS and UTSB.
We use the [CLS] vector as input to a single layer neural network to obtain the probability of the document being relevant.  
We refer to the probability prediction values as UTS and UTSB ranking scores and use the ranking orders they impose to study whether neural models benefit from the body text. 

%UTSB: aether://experiments/ef087c70-dcc9-4a49-8c27-e83bb246c8b2
%UTS:  aether://experiments/a4c96dea-ff1b-432d-b290-c72d8654b28a

%% file: 05-method.tex
For our dataset, we sample 1k queries at random from Bing's search logs, then for each query, we scrape the Bing SERP and collect a total of 12k query-URL pairs. We collect human labels for every query-URL and run ranking experiments with our dataset as the test set. 
For our investigation of when the body text impacts a human/machine assessor, we focus on the query and document properties listed in Table~\ref{tab:features}.

\begin{table*}
  \caption{Query and document features}
  \label{tab:features}
  \begin{tabular}{p{3.6cm}p{9cm}}
    \toprule
    Variable & Description\\
    \midrule
    Performance predictor & Output of a proprietary query performance prediction model ($\in[0,1]$)\\
    Query type: Navigational & Classifier output predicting if the query is navigational (1) or not (0)\\
    Query type: Head/tail & Predicted query popularity ($\in[0 (tail),1 (head)]$) \\
    Query type: Question & If the query is a natural language question ($\in[0 (no),1 (yes)]$)\\
    Lengths & Query, URL, Title, Snippet, and Body lengths in characters\\ % $\in\mathbb{N}$
    \% of query tokens  & The ratio of query tokens that appear in the URL, Title, Snippet, Body\\
  \bottomrule
\end{tabular}
\vspace{-3em}
\end{table*}

%% file: 06-results.tex
\section{Results and Discussions}
\label{sec:result}

\textbf{Impact of body text on human assessors:}
We stipulate that UTS alone is insufficient in cases when human assessors either revised their initial assessment upon seeing the body text (`Revised') or when the body text was needed to confirm their UTS label (`Confirmed'). 
Overall, assessors indicated that UTS alone was insufficient (body text was beneficial) in 48\% of the cases. Of these, `Revised' made up 59\% and `Confirmed' the other 41\%. 
When assessors revised their ratings, they initially overestimated the document's relevance in 54\% of cases (UTS>UTSB) and underestimated it in 46\% of cases (UTS<UTSB). The higher ratio of overestimates could hint at possible SEO manipulation methods succeeding or assessors exhibiting confirmation bias with UTS. 
Using statistical analysis (t-test) to compare the sample means of the query document properties (Table~\ref{tab:features}) across cases where the body text benefited judges or not, we found that the body text was helpful for predictably poor performing, long, not-navigational, tail and question type queries (all stat. sig. p<0.01).

% \begin{table*}
%   \caption{Comparison of the query and document features' sample means against the population means for the sets of judged query-URL pairs where judges did or did not benefit from seeing the body text: all shown values are stat. sig. at p=0.01, except \dag which indicates p=0.05.}
%   \label{tab:samplemeandiffs}
%   \begin{tabular}{c|ccc|ccc}
%     \toprule
%      & Revised & Confirmed & Not needed & UTS > UTSB & UTS < UTSB & UTS = UTSB\\
%     Predicted performance & low & low & high & low & & \\
%     Navigational      & no & no & yes & no & & yes\dag \\
%     Head              & tail & tail & head & tail & & head \\
%     Question          & yes & yes & no & & yes & \\
%     \midrule
%     Query length   & long & long & short & long & & \\
%     URL length     & long\dag &   & short\dag & long & & \\
%     Title length   &  &  & & & & \\
%     Snippet length &  &  & & & short\dag & \\
%     Body length    &  &  & & short & long & \\
%     \midrule
%     \% of query tokens in URL     &  & low & high & low & high & \\
%     \% of query tokens in Title   &  & low & high & low & high\dag & \\
%     \% of query tokens in Snippet &  & low & & & & \\
%     \% of query tokens in Body    &  &  & & low & high & \\
%     \midrule
%     Judging time   & long & long\dag & short & long & long & short \\
%     \bottomrule
% \end{tabular}
% \end{table*}

\textbf{Impact of body text on neural ranker:}
We assume that UTS is insufficient when the UTSB model outperforms the UTS model. We calculate the two models' performance using RBP \citep{rbp} with both the human UTS and UTSB labels as ground-truths. As it can be seen in Table~\ref{tab:rbp}, the UTSB model outperforms the UTS model ($\Delta$RBP>0), where the benefit from body text is more evident at the top ranks ($\Delta$RBP@3>$\Delta$RBP@10). We also see that the ranker learns to make better use of the body text when the training labels also consider the body text (2nd row). Looking at the ratio of queries where the UTSB model outperforms the UTS model (3rd row), we see that there is room for improvement: the percentage of queries that benefit from the body text is just higher than those that body text degrades. Differences in the sample means of the query document properties (Table~\ref{tab:features}) for the improved and degraded queries reveals that improved queries are long, tail, not-navigational and of question type, while degraded queries are short, head and navigational, and the documents long (all stat. sig. p<0.01).

\begin{table}[t]
\begin{minipage}[b]{.45\textwidth}
  \caption{The UTSB model's performance improvement over the UTS model, measured using RBP (on a 100 point scale) and either the UTS or UTSB human labels as ground-truth (GT).}
  \label{tab:rbp}
  \begin{tabular}{lcc}
    \toprule
    & $\Delta$RBP@3 & $\Delta$RBP@10\\
    UTS label GT & 0.165 & 0.071 \\
    UTSB label GT & 0.797 & 0.587 \\
    \% improved/degraded & 33/31 & 45/43 \\
  \bottomrule
\end{tabular}
\end{minipage}
\hfill
\begin{minipage}[b]{.45\textwidth}
  \caption{Reasons when human assessors could not highlight parts of the body text to explain why it was beneficial over the UTS}
  \label{tab:nohighlight}
  \begin{tabular}{lcc}
    \toprule
    & UTS>UTSB & UTS<UTSB\\
    Missing term & 76\% & 12\% \\
    Other & 20\% & 48\% \\
    Video & 4\% & 40\% \\
  \bottomrule
\end{tabular}
\end{minipage}
\end{table}

\textbf{Explanation of body text's impact:}
%We examine under what conditions did the body text impact humans and machines in their respective assessments of a document's relevance. Do humans and neural models react to the additional information in the body text in the same way or does body text impact humans and model differently? %We focus our analysis on the query and document features listed in Table~\ref{tab:features}.
%
We make use of the interpretML framework\footnote{\url{https://interpret.ml/docs/ebm.html}} and train two Explainable Boosting Machine (EBM) glassbox regression models (tree-based, cyclic gradient boosting Generalized Additive Models) \citep{ebm}. For each query-URL pair input, we use the properties listed in Table~\ref{tab:features} as features and construct the target labels as follows:
\vspace{-\topsep}
\begin{itemize}
    \item \textbf{$\Delta$Label}: Target label for the EBM model used to explain human assessors' reaction to seeing the body text, mapped as -1 if UTS>UTSB (UTS label overestimated document relevance), 0 if UTS=UTSB, and 1 if UTS<UTSB (UTS underestimated). %, we first take the difference between the UTSB and UTS labels (UTSB-UTS) for each query-URL pair and then binarize the delta to only consider directionality. Thus, we end up with:% three scores: 
    %-1 if the UTS label overestimated the document's relevance, %(assessors reduced their assessment after seeing the body text: UTS > UTSB), 
    %0 if assessors didn't revise their initial assessment (UTS=UTSB), and +1 if the UTS label underestimated the document's relevance (UTSB > UTS).
    \item \textbf{$\Delta$Rank}: To model the neural rankers' reaction we opt to use the ranking position (rp) since the UTS and UTSB scores are not directly comparable (different trained models) and use -1 if UTS rp < UTSB rp (UTSB model's relevance estimation decreased compared to UTS), 0 if UTS rp = UTSB rp, and 1 if UTS rp > UTSB rp (UTSB's estimate increased compared to UTS). % and scores are also not comparable across queries for a given model). 
    %We sort the results per query using the respective model's UTS or UTSB output scores to derive our rankings, then we take the difference between the UTS and UTSB ranks (UTS-UTSB, i.e., opposite to labels, since a higher rank reflects lower estimated relevance) per query-URL pair and finally we binarize the delta. Thus, -1 means that the UTS model overestimated (ranked a document higher than the UTSB model), 0 means that there was no difference in the rank position, and +1 means that the UTS ranker underestimated the document's relevance.
\end{itemize}

%Figure~\ref{fig:ebm} 
Table~\ref{tab:ebm} shows the EBM models' top 5 feature importance scores for human and machine assessors, telling us which of the query and document properties explain the delta observed in the human assessors' UTS and UTSB labels ($\Delta$Label) and the neural models' UTS and UTSB based rankings ($\Delta$Rank), respectively. %\footnote{Note that we injected generated random numbers (randnum) as an additional feature as a way of forming a baseline where features start to flatten out.}. 
We can see that a change in labels or rankings is explained by very different factors: body length is the only common factor in the top 5. %humans react to the body text under different conditions than the neural models. 
The top explanation of change in the humans' UTS vs UTSB assessments is whether the query is phrased as a question, while the top reason for the ranker is the ratio of query tokens that are present in the body text. %The top 3 most important query or document properties that explain when human assessors stick with or revise their initial UTS assessments are whether the query is a Question, the length of the document (Body length), and the output of our Performance predictor. %, the ration of query tokens that appear in the document title (\%QueryWords in Tokenized Title) and the Query length. 
%On the other hand, the top most important features that explain when the neural ranker re-ranks documents after 'seeing' the body text are the ratio of query tokens that appear in the body text (\%QueryWords in Tokenized Body), the Snippet and Title lengths. %and Body lengths and the ratio of query tokens that appear in the Snippet (\%QueryWords in Tokenized Snippet).  
%
%The differences in the feature importances highlight that humans and neural models process the information present in the UTS and body differently and benefit from the body in different ways. %It is possible that assessors are sensitive to different aspects or that miss some information due to layout, while neural models 
%TODO: add more discussion here!

% \begin{figure}[t]
%   \centering
%   \includegraphics[width=1\columnwidth]{fig/bodyvec-cikm2021-ebm.JPG}
%   \caption{The EBM models' feature importance scores for the human (left) and machine (right) assessors, explaining their reaction to the body text.}
%   \label{fig:ebm}
% \end{figure}

\begin{table}[t]
%\begin{minipage}[b]{.45\textwidth}
  \caption{The EBM models' top 5 feature importance scores for human and machine assessors, explaining the delta observed in the human assessors' UTS and UTSB labels ($\Delta$Label) and the neural models' UTS and UTSB based rankings ($\Delta$Rank), respectively.}
  \label{tab:ebm}
  \begin{tabular}{ll}
    \toprule
    \textbf{$\Delta$Label} (UTSB label - UTS label) & \textbf{$\Delta$Rank} (UTS rp - UTSB rp)\\
    Question (0.2825) & \%QueryWords in Tokenized Body (0.2858) \\
    Body length (0.2434) & Snippet length (0.2831) \\
    Performance predictor (0.2418) & Title length (0.2478) \\
    \%QueryWords in Tokenized Title (0.2218) & Body length (0.1658)\\
    Query length (0.2141) & \%QueryWords in Tokenized Snippet (0.1459)\\
  \bottomrule
\end{tabular}
%\end{minipage}
\end{table}

\begin{figure}[t]
  \centering
  \includegraphics[width=0.92\columnwidth]{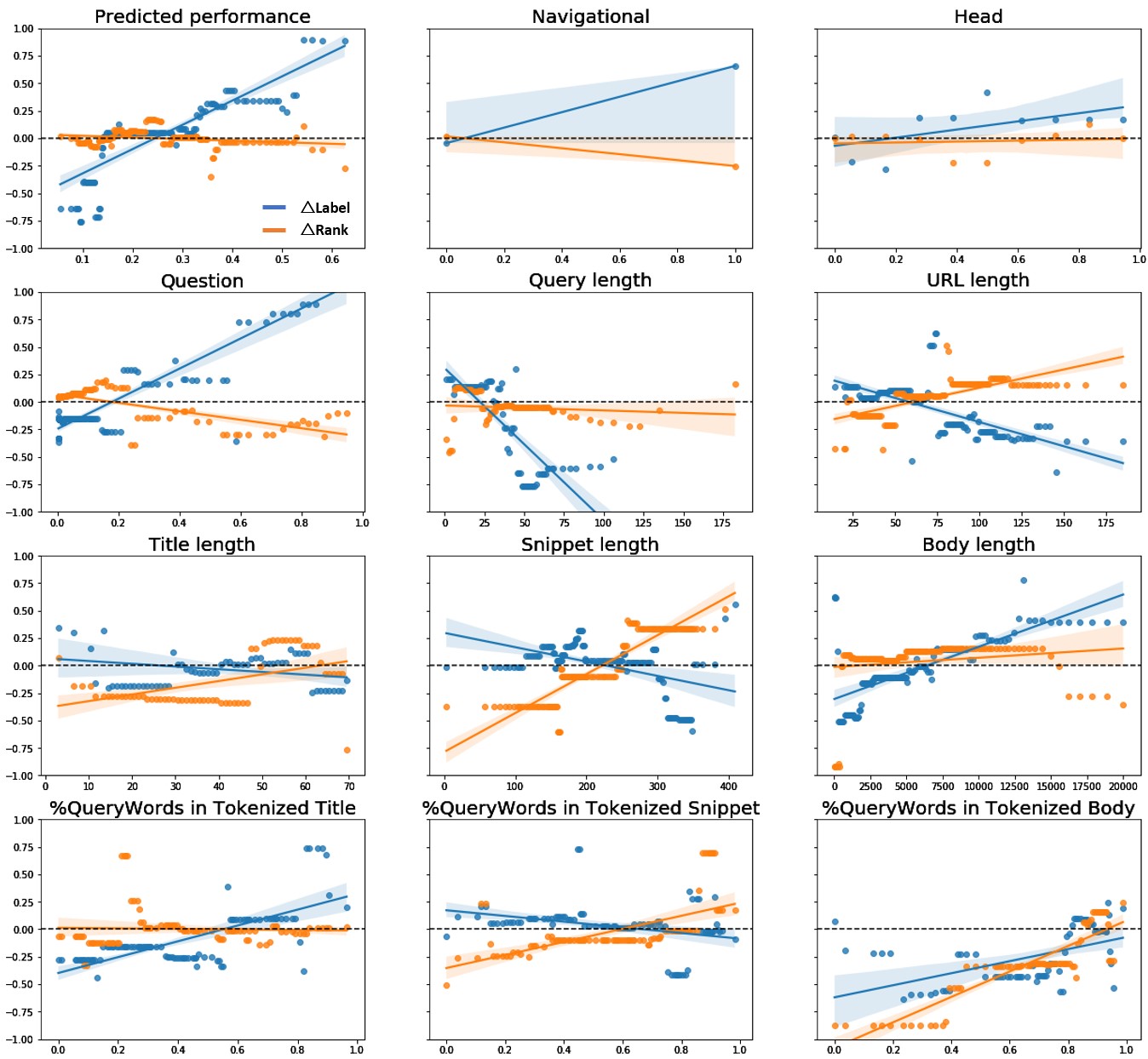}
  \caption{EBM's learnt feature functions for each query and document feature, explaining the $\Delta$ changes: $y>0$ means that `seeing' the body text led to an increase in the relevance estimate compared to UTS}
  \label{fig:ebmfeatures}
\end{figure}

To examine how $\Delta$Label and $\Delta$Rank change with each feature, in Figure~\ref{fig:ebmfeatures}, we plot EBM's learnt per-feature functions. Each plot shows how a given feature contributes to the model’s prediction. For example, the Query length plot shows that for short queries, human assessors (blue line) are more likely to underestimate (y>0) the document's relevance based on UTS alone, while for long queries, they tend to overestimate (y<0). The neural model (orange line) shows a similar but more subtle trend: for short queries, body text increases the ranker's relevance estimate over the UTS estimate, while for long queries the predicted relevance decreases with body text. The Question plot shows that humans tend to underestimate the document's relevance when the query is more likely to be a question. This indicates that document summaries fail to convince searchers that the document answers their question. The ranker's predicted relevance, however, decreases with body text for question type queries. Looking at the Snippet length plot, we see that the neural model is more likely to decrease its estimate of the document's relevance with body text when snippets are short, but increase it for long snippets. This suggests that when snippets include more context, the ranker is more likely to see these as evidence of irrelevance, which is diminished when body text is added. Snippet length has the opposite impact on humans: the longer the snippet, the more likely they overestimate the result's relevance. Overall, we see very little similarities (parallel trends) in the human vs ranker feature plots, indicating that humans and machines react to body text in fundamentally different ways.
%
%Considering feature weights around 0 as cases where query-biased summaries were deemed by humans or neural model as sufficient and feature weights further away from 0 as cases where UTS was insufficient, 
Human assessors are more likely to overestimate relevance from UTS for long, tail, and not-navigational queries, and underestimate when the query is head, navigational or a question. They also overestimate for long snippets and short documents, and underestimate for long documents and short snippet. Unlike humans, the neural model results in more near-flat plots: the most impact is seen for document (rather than query) properties, e.g., Snippet length and ratio of query tokens in the snippet and body.  

% We can also see that our performance predictor predicted low retrieval performance for the cases where judges benefited from the body text. Looking at when judges initially over- or underestimated the document's relevance, we can see that they were more likely to overestimate for long, tail, and not-navigational queries, but tended to underestimate when the query was a question. We can also see that shorter web pages were likely to be overestimated, while longer pages tended to be underestimated.

\textbf{Additional considerations:} %An aspect that can influence human assessors but is not considered by the neural model is the impact of non-textual elements, such as video.
When assessors revised their relevance assessment but were unable to highlight parts of the body text to explain the change (in 72\% of overestimates and 22\% of underestimates), they were asked to indicate a reason. Table~\ref{tab:nohighlight} shows that the absence of query terms in the document was the main reason for overestimates without highlighted text (76\%). This suggests that informing users of missing query terms on the SERP is a helpful strategy. On the other hand, a major reason when assessors underestimated a document was when video (or other non-textual content) was present on the page (40\%) - an aspect that was not considered by the neural model.

%% file: 07-conclusion.tex
\section{Conclusions}
\label{sec:conclusion}

We studied when human and machine assessors benefit from the full text of the document to estimate its relevance. We showed that both humans and BERT style models benefit from the body text in similar cases (long, not navigational, tail and question type queries), but that full text impacts their relevance assessments in different ways (e.g., full text increases humans' relevance estimates but decreases the ranker's). In addition, we observe differences in the properties of queries where the BERT model's performance improves or degrades with the full text, e.g., performance degrades for navigational queries ($\Delta$RBP@3 of -1.07). 
This indicates that more work is necessary on BERT style models when considering full text as input or that different types of queries (e.g., head v tail) require models to be optimized differently. While our findings are a function of the query-biased summaries, the observed differences in human and model reactions to additional information indicate that different mechanisms are needed for human vs machine inputs. 